# Millimeter Wave and Terahertz Spectra and Global Fit of Torsion-Rotation Transitions in the Ground, First and Second Excited Torsional States of $^{13}CH_3OH$ Methanol

Li-Hong Xu[a], R.M. Lees, Yun Hao

Centre for Laser, Atomic and Molecular Sciences (CLAMS), Department of Physics,

University of New Brunswick, Saint John, N.B., Canada E2L 4L5

H.S.P. Müller, C.P. Endres, F. Lewen, S. Schlemmer

I. Physikalisches Institut, Universität zu Köln, 50937 Köln, Germany

K.M. Menten

MPI für Radioastronomie, 53121 Bonn, Germany

Suggested Running Head:

Global Fit of Torsion-Rotation Transitions for $^{13}CH_3OH$

Correspondence to be sent to:

Dr. Li-Hong Xu

Centre for Laser Applications and Molecular Science

Department of Physics

University of New Brunswick

Saint John, N.B.  Canada  E2L 4L5

Phone:   506-648-5632

Fax:       506-648-5650

E-Mail:  lxu@unb.ca

# ABSTRACT

Methanol is observed in a wide range of astrophysical sources throughout the universe, and comprehensive databases of the millimeter and THz spectra of $CH_3OH$ and its principal isotopologues represent important tools for the astronomical community. A previous combined analysis of microwave and millimeter wave spectra of $^{13}CH_3OH$ together with Fourier transform far-infrared spectra was limited to the first two torsional states, $v_t = 0$ and 1, for $J$ values up to 20. The limits on frequency and quantum number coverage have recently been extended by new millimeter and THz measurements on several different spectrometers in the Cologne laboratory in the frequency windows 34-70 GHz, 75-120 GHz, 240-340 GHz, 360-450 GHz and 1.12-1.50 THz. With the new data, the global treatment has now been expanded to include the first three torsional states for $J$ values up to 30. The current $^{13}CH_3OH$ data set contains about 2,300 microwave, millimeter-wave, sub-millimeter and THz lines and about 17,100 Fourier-transform far-infrared lines, representing the most recent available information in the quantum number ranges $J \leq 30$, $K \leq 13$ and $v_t \leq 2$. The transitions have been successfully fitted to within the assigned measurement uncertainties of ±50 kHz for most of the frequency-measured (i.e. MW, MMW, Sub-MMW, THz) lines and ±6 MHz for the FIR lines. A convergent global fit was achieved using 103 adjustable parameters to reach an overall weighted standard deviation of 1.37. Our new C-13 methanol database is improved substantially compared to the existing one (Li-Hong Xu, F.J. Lovas, J. Phys. Chem. Ref. Data 26 (1997) 17-156), and will be available in the Cologne Database for Molecular Spectroscopy, CDMS (http://www.astro.uni-koeln.de/cdms/), in support of astronomical studies associated with results from HIFI (Heterodyne Instrument for the Far-Infrared) on the Herschel Space Observatory and new observations from SOFIA (Stratospheric Observatory For Infrared Astronomy) and ALMA (Atacama Large Millimeter/Submillimeter Array).

# I. INTRODUCTION

Methanol is an important diagnostic tool of interstellar clouds [1-4]. It is so abundant in some astronomical sources that it accounted, together with $SO_2$, for half of the integrated emission in the 607–725 GHz molecular line survey of the prominent Orion KL star forming region [5] and is equally rich in the central galactic source Sgr B2(N) up into the THz region [6]. Therefore, it is not surprising that $CH_3OH$ was among the earliest molecules detected in space as the fifth polyatomic species [7], while its $^{13}C$ isotopologue was detected only a few years later [8]. Numerous lines of the $^{13}C$ species have now been seen in the surveys, providing important information for deriving proper column densities of the main isotopologue because lines of the latter frequently become saturated in dense regions (see [4,9,10], for example). The $^{12}C/^{13}C$ ratio, 89 in the Solar system, differs considerably throughout the Galaxy from ~20 near the centre to ~60 in the Solar neighbourhood to possibly more than 100 in the outer regions of the Milky Way, and it decreases in time [11]. Hence, the $^{12}C/^{13}C$ ratio in methanol may, in turn, provide significant clues about the processes producing this molecule in the interstellar medium [12], as well as about routes for formation of more complex molecules from this abundant species [13,14].

As mentioned above, methanol and its isotopologues are among the most prolific contributors to wide-band radio astronomical survey spectra from interstellar space [4-6,9,10,15-17]. With the increased sensitivity, resolution and frequency coverage of recent telescopes and observatories, such as HIFI, ALMA and SOFIA, it is certain that even larger numbers of new and unknown lines will be observed than in the past, challenging the astronomers. A substantial proportion of such unknown lines will likely arise from high rotational and torsional states of known molecules, with the methanol isotopologues $CH_3OH$, $^{13}CH_3OH$, $CH_3OD$, etc., being major contributors. For this reason, the database for the primary $CH_3OH$ species was substantially improved and extended in 2008 to include the first three torsional states ($v_t$ = 0, 1 and 2) for $J$ rotational quantum number up to 45, based on a fit to $J_{max}$ = 30 [18].

The next most abundant isotopic species of methanol is $^{13}$CH$_3$OH, for which the existing database dates back to 1997 [19] when a line list up to 1 THz was generated based on a fit of the first two torsional states ($v_t$ = 0 and 1) for $J$ up to 20 [20]. However, the absence of measured or predicted data above 1 THz is already being felt by astronomers in the analysis of HIFI spectra [4], and is particularly important since methanol is the major weed molecule with line density persisting significantly up into the THz region [15]. Although further experimental information on higher torsional and rotational states is available from reported MMW [21] and Fourier transform far-infrared (FTFIR) studies [22], a substantial number of lines in the latter are blended at the FTFIR resolution while the measurement accuracy for unblended lines is of order 0.0002 cm$^{-1}$ or 6 MHz. This may not be sufficient to overcome the problem of line confusion that is likely to arise with the enhanced capabilities of the latest astronomical facilities. Thus, to expand the database of precise frequency information for the astronomical community, we have extended measurements in the current work up to the Sub-MMW and THz regions and have then carried out a global analysis of the MW, MMW, Sub-MMW, THz and FIR data for $^{13}$CH$_3$OH in the first three torsional states ($v_t$ = 0, 1 and 2) of the ground vibrational state. The newly measured and previously reported transitions for C-13 methanol involving $J$ quantum numbers up to 30 have been combined in our analysis using an improved version of the BELGI global fitting program [23,24] based on a one-dimensional torsion-rotation Hamiltonian [25] expanded to include a selection of new terms up to 10$^{th}$ order in the torsion-rotation operators.

## II. MEASUREMENTS

Millimetre wave (MMW) and terahertz spectra of $^{13}$CH$_3$OH (99 % $^{13}$C, Sigma Aldrich) were recorded at Universität zu Köln. All measurements were carried out at room temperature and pressures around 2 Pa. Single-path absorption cells were used for all spectrometers, consisting of large 10-cm i.d. Pyrex tubes with Teflon lenses to focus the radiation in and out. Frequency modulation was used in all instances with demodulation at 2*f*, producing absorption line profiles close to the second derivative of a Gaussian.

The frequency range 34–70 GHz was covered by a commercial synthesizer (Agilent E8257D) as source and a home-built Schottky diode as detector, using a 7-m absorption cell. The 75-120 GHz region was covered with essentially the same setup, except that the synthesizer was used to drive a frequency tripler (Virginia Diode, Inc.). Measurements between 240 and 340 GHz as well as between 370 and 450 GHz were carried out with the Cologne Terahertz Spectrometer (CTS) [26] employing phase-locked backward wave oscillators OB30 and OB32, respectively, as sources, a liquid He-cooled InSb hot-electron bolometer (QMC Instruments Ltd.) as detector, and a 4-m absorption cell.

The THz spectrometer, described in our parallel study of $CH_3SH$ [27], employed a MW synthesizer driving a multiplier chain (Virginia Diodes, Inc.) containing either a doubler or tripler amplifier followed by a sequence of 2 doublers and 2 triplers to provide output at the 72$^{nd}$ harmonic from approximately 1.1-1.34 THz on low band and the 108$^{th}$ harmonic from 1.33-1.50 THz on high band. Again, a 4-m absorption cell was used with a He-cooled QMC bolometer detector.

In general, the lines on all spectrometers were observed with very symmetric line shapes, permitting accurate determination of the center frequencies. Our operating and analysis procedure [27] yielded line positions with measurement accuracy believed to be better than 50 kHz for the great majority of the observed transitions.

In the measurement campaign, we employed two lists of predicted frequencies to guide the search. The first was the compilation of calculated line positions with uncertainties and line strengths similar to data available in Table 5 of Ref. [19], but with quantum numbers increased to $v_{tmax} = 2$ and $J_{max} = 45$ using parameters from global fits expanded over Ref. [20] to include MMW and FTFIR data for $v_t = 2$ [21,22]. The second was generated by taking differences between the $^{13}CH_3OH$ energy term values reported in the Supplementary Data to Ref. [22] in order to calculate the frequencies of all conventionally allowed transitions up to 2 THz with an expected FIR measurement accuracy of order ±10 MHz. This allowed the measurements to be carried out in a targeted fashion either line-by-line or in short sweeps, with the synthesizer set to scan a small region around the predicted frequency of each desired transition. In the later stages with the THz spectrometer, broad scans were also conducted covering several GHz in order to explore long *Q*-branch *J*-progressions or to cover a significant portion of a wide *a*-type *J* multiplet. Initially, since the majority of previously reported MMW measurements [21] involved $\Delta K = 0$ *a*-type lines that are relatively insensitive to many of the important torsional molecular parameters, our emphasis was on *b*-type transitions for which there can be substantial changes in torsional energy. As more such data were accumulated, the global fitting calculations and predictions were refined, and a further focus in the measurements then became to locate those transitions with larger predicted uncertainties that would supply the greatest information content in tying down the parameters.

So far, from 34.7 to 127.6 GHz we have assigned nearly 150 new transitions, as well as remeasuring several previously reported lines to check our agreement. From 244.3 to 453.3 GHz we have assigned around 220 new lines, and from 1.12 to 1.50 THz we have over 1100 new assignments. The new data cover energy levels with a wide range of quantum numbers, and include a number of "forbidden" lines with $|\Delta K| > 1$ that were predicted from the global model to arise with significant intensity due to strong asymmetry mixing among the low-*K* levels of *E* torsional symmetry. Several lines were also observed from each of three $v_t = 2 \leftarrow 1$ sub-bands that are particularly low-lying due to substantial cancellation of rotational and torsional energy and thus are especially sensitive to the molecular parameters.

Figure 1 shows part of one of the broad-band THz scans, displaying the two strong *Q* branches corresponding to the $K = 7 \leftarrow 6$ and $-7 \leftarrow -6$ $v_t = 0$ transitions of *E* torsional symmetry (originally located with unresolved origins in the FTFIR work [22]). The $-7 \leftarrow -6$ *E Q* branch on the right shades toward lower frequency with steadily increasing spacing between consecutive *J* transitions, whereas the $7 \leftarrow 6$ *E* branch on the left shades to lower frequency initially but then converges to form a *Q*-branch head and turns around. This rather dramatic difference in behaviour for two similar transitions arises from subtle differences in the contributions of molecular asymmetry shifts and torsion-rotation distortion terms to the effective *B* rotational constants and *D* centrifugal distortion parameters for the relevant states. With the customary expansion of the energy levels of a sub-state as a series in powers of $J(J + 1)$, the frequency of a *Q*-branch transition can be written as

$$\nu_Q(J) = \Delta E_o + \Delta B_{eff} J(J + 1) - \Delta D_{eff} J^2(J + 1)^2 + \ldots \tag{1}$$

where $\Delta E_o$ is the difference between upper and lower sub-state origins, $\Delta B_{eff} = B' - B''$ is the difference between effective *B*-values, and $\Delta D_{eff} = D' - D''$ is the difference between effective centrifugal constants. For medium to high *K* transitions, $\Delta B_{eff}$ is normally negative while $\Delta D_{eff}$ is sensitive to the asymmetry shifts of the levels so can have either sign. From Eq. (1), the frequency difference between successive lines in a *Q*-branch is given by

$$\Delta \nu_Q(J) = \nu_Q(J) - \nu_Q(J - 1) = 2\Delta B_{eff} J - 4\Delta D_{eff} J^3 \tag{2}$$

Thus, if $\Delta B_{eff}$ and $\Delta D_{eff}$ have opposite signs, the *Q*-branch spacing will increase monotonically with *J*, whereas if they have the same sign $\Delta \nu_Q(J)$ will go to zero and a *Q*-branch head will form for a value of *J* given by

$$J_{Q\text{-head}} = (\Delta B_{eff} / 2\Delta D_{eff})^{1/2} \tag{3}$$

In the present case, calculation with a simplified torsion-rotation model evaluating the asymmetry shifts by second-order perturbation theory gives $\Delta B_{eff}$ and $\Delta D_{eff}$ values of –2.94 and –0.00205 MHz for the 7 ← 6 *E* transition, and –1.56 and +0.0091 MHz for –7 ← –6, respectively. Therefore, the line spacing for the latter will keep increasing, as observed, while that for the former should go to zero at $J = 26$, reasonably consistent with the observed value of $J = 25$ given the approximate nature of the model and the neglect of higher-order terms.

The third small *Q* branch in the middle of Fig. 1, which appeared earlier only as a broad unresolved absorption feature in the FTFIR spectrum, has not so far been identified. It is not present in either of our lists of predictions so represents an intriguing assignment challenge for the future.

Figure 2 illustrates a segment of the $J = 25 \leftarrow 24$ *a*-type $\Delta K = 0$ multiplet for which the *K* rotational quantum numbers do not change in a transition. Since the rotational constants are relatively similar for the different *K* states in the *J*-multiplets, the lines tend to cluster in dense groups and are often overlapped. Many of the transitions in Fig. 2 are assigned in the Supplementary Data to Ref. [22], but a number of reasonably strong features are missing there and numerous lines reported as blended are clearly resolved in Fig. 2, which would be important for astronomical applications.

Figures 1 and 2 illustrate the particular value of methanol as a probe of excitation conditions in astronomical sources. For the $Q$ branches of the form of Fig. 1, a wide range of $J$ states is covered for given $K$, $v_t$ and torsional symmetry in a compact group of lines. Conversely, in a $J$-multiplet such as seen in Fig. 2, the $J$-values stay the same but a wide range of $K$ and $v_t$ states is observed, again in a tight group of transitions. Thus, both types of spectral structure enable astronomical observations covering a broad range of excitation energies within a narrow frequency window over which the receiver behaviour should be uniform, ensuring accurate relative intensity determination. For the important classes of high-density regions in which the line intensities for the C-12 parent are strongly affected by saturation, such observations for optically-thin line groupings of the C-13 species will then become especially useful in giving well-determined intensity comparisons and thereby permitting reliable deduction of source temperatures and densities.

### III. GLOBAL MODELING

#### 1. The $v_t = 0, 1$ and 2 Data Set

The current data set builds on the work of Ref. [20], with extended MMW [21] and FTFIR [22] information, addition of further MW, MMW and Sub-MMW transitions from the NIST MW data center compilation [28] and new measurements from the Cologne laboratory as described in Sec. II above. Most frequency-measured MW, MMW, Sub-MMW and THz lines were assigned a ±50 kHz uncertainty with the exception of unresolved or barely resolved $K$-doublet or overlapped transitions which were assigned ±100 or ±200 kHz uncertainties. FTFIR lines were assigned uncertainties of ±0.0002 or ±0.00035 cm$^{-1}$ (±6 MHz or ±10.5 MHz) for transitions originating from $v_t = 0$ or $v_t = 1$ and 2 levels, respectively. The final data set contained a total of ~19,400 lines including ~2300 MW, MMW, Sub-MMW and THz lines and ~17,100 FTFIR lines.

In a small number of cases, assigned frequency-measured lines were weighted to zero and not included in the fit if the ($\nu_{obs} - \nu_{calc}$) residuals were unreasonably high, if the lines were clearly blended and/or had multiple assignments, or if the lines belonged to a sub-band that appeared to be perturbed with systematically large residuals of several MHz. Similarly, several FIR sub-bands with large and systematic residuals 0f 0.002 cm$^{-1}$ or more were excluded from the fit. As seen from the energy level plot for the analogous O-18 species in Fig. 4 of Ref. [29], near-degeneracies can arise between $v_t = 3$ and high-lying $v_t = 2$ and levels and also between high-$K$ {($K$, $v_t$), ($K$-2, $v_t$+1), ($K$-4, $v_t$+2)} level triplets. We believe these are the likely source of the perturbations, as they involve interaction partner states not included in our data set at present. For the frequency-measured lines, the maximum $K$ values included in the fit were: 13$A$, –10$E$ and 12$E$ for $v_t = 0$; 11$A$, –10$E$ and 11$E$ for $v_t = 1$; and 8$A$, –7$E$ and 7$E$ for $v_t = 2$. The affected FIR sub-bands involved principally the higher-lying $v_t = 2$ levels, including the 2$A$, 5$A$, 9$A$, 12$A$, –10$E$, 4$E$ and 11$E$ states, that showed $J$-independent downshifts of up to -0.01 cm$^{-1}$ from positions calculated from the fitted model.

### 2. Over-All Fit Results and Parameters

The statistics of our final data set are outlined in Table 1 to illustrate the range of the data and the relative quality of the fit over the different torsional states. As with other multi-parameter least-squares fits, the questions of when to stop trials and how to choose the best set of final parameters are open to debate. In the present case, we relied on experience gained through considerable trial and error here as well as from parallel studies of other similar species such as CH$_3$OH, CH$_3$$^{18}$OH and CH$_3$SH with similar quantum state coverage. We were ultimately able to achieve a satisfactory convergent global fit with an overall unitless weighted standard deviation of 1.37 using 103 adjusted parameters. These are reported in Table 2 and include the traditional rotational and torsional constants as well as cross terms up to 8$^{th}$ order. For comparison, the parameters of the CH$_3$OH parent species are also listed in Table 2. Most of the parameters are reasonably consistent in magnitude and sign for the two species. Note that the changes in parameters from the previous $^{13}$CH$_3$OH $v_t$ = 0-1 global fit values [19,20] are substantially larger than the quoted rms parameter errors in the earlier work, by factors of 372 for $V_3$ and 11 for $F$, for example. The changes are similar to those found for the C-12 parent [18,19], and reflect very high correlations that are encountered among the torsional parameters in this type of problem. Thus, when a new $v_t$ state is added to the fit bringing significant further information on the torsional potential, parameters such as $V_3$ and $F$ tend to readjust by amounts much greater than just the statistical fitting uncertainties.

Also, as was again found earlier in the fitting for the parent species [18], the rms deviation for the MW lines in Table 1 is a good deal larger than the estimated measurement uncertainty of 50 kHz, with the $v_t$ = 0 lines being fitted to 111 kHz and the $v_t$ = 1 and $v_t$ = 2 lines being somewhat worse at a little over 150 kHz. Blending and overlapping from underlying weak lines for a number of transitions would certainly contribute to this, but we believe that the principal source still lies in the modeling. A significant feature of the fit is that the residuals for many of the sub-bands are not random but show systematic shifts from the calculations, particularly at high $J$ and $K$ values, implying that the model is not capturing all of the torsion-rotation energy dependence of the data. This might indicate problems with the form of the model as a power-series expansion of torsion-rotation operators, but also could be due to the restricted range of states including in the fitting. For instance, if anharmonic resonances between the CO-stretching vibrational state and the $v_t$ = 3 and 4 torsional levels can produce perturbations at the cm$^{-1}$ level [30], then shifts in the MHz range might well be expected for the $v_t$ = 0 to 2 states down below. To properly diagnose this problem, the fitting domain should really be expanded to include the higher torsional levels and the low-lying vibrational states, which represents a rather substantial undertaking for the future.

## 3.  $^{13}$CH$_3$OH line list

Employing the parameters of Table 2, we have generated a line list up to $J_{max\_calc} = 40$, extrapolating from our fit with $J_{max\_fit} = 30$ but not extrapolating in the $v_t$ quantum number. With the extensive data set and wide range of quantum states included in the fit, we believe the accuracies of the predicted lines will range from ~50 kHz to ~6 MHz, depending on whether frequency-measured or Fourier-transform FIR data were fitted in the original data set. The uncertainties in the calculated frequencies will increase somewhat as $J$ rises above the $J_{max\_fit}$ upper limit listed. Our results will be made available in CDMS, the Cologne Database for Molecular Spectroscopy [33,34], in order to be accessible to all user communities.

The line strength and uncertainty calculations for the line list were similar to the schemes described in Ref. [19] using permanent dipole moment components $\mu_a = 0.899$ D and $\mu_b = -1.44$ D in the Internal Axis Method (IAM) system. The uncertainties quoted are twice the standard deviations from the least-squares analysis (i.e. 95 percent confidence levels or expanded uncertainty with coverage factor $k = 2$) and were estimated from the variance-covariance matrix as described in Ref. [32].

## IV.  DISCUSSION AND CONCLUSIONS

The present work represents a significant extension to the information base of precisely measured or predicted transitions of the $^{13}$CH$_3$OH isotopologue of methanol. The dataset, to be deposited in CDMS, the Cologne Database for Molecular Spectroscopy [33,34], should be an important asset for the astronomical community in the analysis of sensitive spectral observations from interstellar and protostellar sources in our galaxy and the wider universe.

The new frequency measurements carried out in this work cover several spectral bands from the millimeter to the THz regions, and in conjunction with previous results provide a strong anchor and rigorous test for the global fitting analysis. Particular attention was paid in the measurements to cover transitions that had a significant predicted uncertainty based on the previous global parameters in order to insure maximum information content for the new global fit. In this fit, an extended model was used that includes a selection of torsion-rotation terms up to the 8$^{th}$ order that were found to make determinable contributions. Comparison of the $^{13}CH_3OH$ parameters with those found previously for the main $^{12}CH_3OH$ parent [18] shows quite good consistency between results for the two species, with a few exceptions among the higher order terms. These deviations might be a consequence of the very much larger data set for the main isotopologue, or to subtle changes in perturbation patterns with the shifting of levels within the energy manifold due to the isotopic substitution.

The current predictions are particularly useful for astronomical observations at higher frequencies, e.g., for those which have been carried out with HIFI on the *Herschel* satellite [35] or for those which have been or will be carried out with the Stratospheric Observatory For Infrared Astronomy (SOFIA) [36], in particular with the German REceiver At Terahertz frequencies (GREAT) [37]. They are even more important for studying higher rotationally or vibrationally excited states, e.g., with the Atacama Large Millimeter Array (ALMA) or other radio telescope arrays.

## ACKNOWLEDGMENTS

L.H.X. and R.M.L. gratefully acknowledge financial support of this research by the Natural Sciences and Engineering Research Council of Canada and express their sincere gratitude for the hospitality and support received during their visits to the Cologne laboratory. The work in Köln has been supported by the Deutsche Forschungsgemeinschaft (DFG) through the collaborative research grant SFB 956 (project B3) and the preceding DFG project SCHL 341/5-1. HSPM is very grateful to the Bundesministerium für Bildung und Forschung (BMBF) for initial support through project FKZ 50OF0901 (ICC HIFI *Herschel*) aimed at maintaining the Cologne Database for Molecular Spectroscopy, CDMS. This support has been administered by the Deutsches Zentrum für Luft- und Raumfahrt (DLR).

## LIST OF TABLES

Table 1   Statistics of the data set for the torsion-rotation global fit[a] to $v_t = 0, 1, 2$ torsional states of $^{13}CH_3OH$

Table 2   The 103 torsion-rotation parameters (in cm$^{-1}$) in the global fit of $v_t = 0, 1$ and 2 torsional states of $^{13}CH_3OH$ methanol and comparison with the parent species

**FIGURE CAPTIONS**

Fig.1   Section of the THz spectrum of $^{13}CH_3OH$ illustrating the $K = 7 \leftarrow 6$ and $-7 \leftarrow -6$ $v_t = 0$ $Q$ branches of $E$ torsional symmetry. Several other strong $v_t = 0$ $b$-type transitions are also seen, labelled in $J_K$ notation, plus one $v_t = 1$ line. The small $Q$ branch in the centre of the picture so far is unidentified.

Fig.2   Part of the dense $J_K = 25_K \leftarrow 24_K$ $a$-type $J$-multiplet in the THz spectrum of $^{13}CH_3OH$. Transitions are labelled with the $K$ value and the torsional $A$ or $E$ symmetry; $v_t = 0$ lines (labelled in blue) have no superscript on $K$ while $v_t = 1$ lines (labelled in red) have superscript 1. The $J_K = 21_0 \leftarrow 20_1$ $E$ $v_t = 1$ $b$-type line (labelled in pink) is also seen.

Table 1
Statistics of the data set for the torsion-rotation global fit[a] to $v_t = 0, 1, 2$ torsional states of $^{13}CH_3OH$

| MW[b] | RMS[c] | | # data[d] | FTFIR[b] | Uncertainties[e] | | RMS[c] | | # data[d] |
|---|---|---|---|---|---|---|---|---|---|
| | Unitless | MHz | | | cm$^{-1}$ | MHz | Unitless | cm$^{-1}$ | |
| All MW lines | 2.41 | 0.134 | 2310 | All FTFIR lines | | | 1.16 | 0.00035 | 17093 |
| $v_t = 0 \leftarrow 0$ | 1.92 | 0.111 | 1242 | $v_t = 0 \leftarrow 0$ | 0.0002 | 6 | 0.91 | 0.00018 | 3186 |
| $v_t = 1 \leftarrow 0$ | | | | $v_t = 1 \leftarrow 0$ | 0.0002 | 6 | 1.00 | 0.00020 | 5728 |
| $v_t = 1 \leftarrow 1$ | 2.90 | 0.157 | 754 | $v_t = 2 \leftarrow 0$ | 0.0002 | 6 | 1.44 | 0.00050 | 1488 |
| $v_t = 2 \leftarrow 1$ | | | | $v_t = 1 \leftarrow 1$ | 0.00035 | 10.5 | 0.60 | 0.00021 | 2617 |
| $v_t = 2 \leftarrow 2$ | 2.81 | 0.154 | 314 | $v_t = 2 \leftarrow 1$ | 0.00035 | 10.5 | 1.80 | 0.00063 | 2862 |
| | | | | $v_t = 2 \leftarrow 2$ | 0.00035 | 10.5 | 1.08 | 0.00038 | 1212 |
| Uncertainty[e] in MHz | | | | Uncertainty[e] in cm$^{-1}$ | | | | | |
| 0.050 | 0.123 | | 2180 | 0.00020 | | | | 0.00019 | 8891 |
| 0.100 | 0.156 | | 84 | 0.00035 | | | | 0.00047 | 8199 |
| 0.200 | 0.268 | | 28 | | | | | | |
| 1.000 | 0.506 | | 18 | | | | | | |

_a The overall unitless standard deviation for this fit of 19 403 data to 103 parameters is 1.37.
b The microwave (MW) and Fourier transform far infrared (FTFIR) transitions are grouped first by torsional quantum number $v_t$, and then by their assigned uncertainties in the fit. Weights used for all lines in the fit are 1/(uncertainty)$^2$.
c Weighted (unitless) and unweighted (in MHz or cm$^{-1}$) root-mean-square residuals from the global fit.
d The number of transitions in each category included in the least squares fit.
e Measurement uncertainties assigned to the various types of transitions (type B, k=1 [32]).

Table 2

The 103 torsion-rotation parameters (in cm$^{-1}$) in the global fit of $v_t = 0$, 1 and 2 torsional states of $^{13}CH_3OH$ methanol and comparison with the parent species

| Term Order $\{nlm\}$[a] | Operator[b] | Parameter[b] In Program | Literature[c] | $^{13}CH_3OH$ $v_t = 0$, 1 and 2 (Present Work) | $CH_3OH$ [d] (Ref. 18) |
|---|---|---|---|---|---|
| {220} | $P_\gamma^2$ | FPARA | F | 27.64201624(70) | 27.64684641(28) |
|  | $(1-\cos 3\gamma)/2$ | V3 | $V_3$ | 373.741301(27) | 373.554746(12) |
| {211} | $P_\gamma P_a$ | RHORHO | $\rho^e$ | 0.8101648121(45) | 0.8102062230(37)[e] |
| {202} | $P_a^2$ | OA | A | 4.2538428(51) | 4.2537233(71) |
|  | $P_b^2$ | B | B | 0.8034196(51) | 0.8236523(70) |
|  | $P_c^2$ | C | C | 0.7737925(50) | 0.7925575(71) |
|  | $\{P_a,P_b\}$ | DAB | $D_{ab}$ | -0.0043475(87) | -0.0038095(38) |
| {440} | $P_\gamma^4$ | AK4 | $F_m$ ($k_4$) | -8.969218(80)x10$^{-3}$ | -8.976763(48)x10$^{-3}$ |
|  | $(1-\cos 6\gamma)/2$ | V6 | $V_6$ | -1.33473(17) | -1.319650(85) |
| {431} | $P_\gamma^3 P_a$ | AK3 | $\rho_m$ ($k_3$) | -3.500229(26)x10$^{-2}$ | -3.504714(14)x10$^{-2}$ |
| {422} | $P_\gamma^2 P^2$ | GV | $F_J$ ($G_v$) | -1.394(22)x10$^{-4}$ | -1.373(31)x10$^{-4}$ |
|  | $P_\gamma^2 P_a^2$ | AK2 | $F_K$ ($k_2$) | -5.179988(32)x10$^{-2}$ | -5.188031(18)x10$^{-2}$ |
|  | $P_\gamma^2\{P_a,P_b\}$ | DELTA | $F_{ab}$ ($\Delta_{ab}$) | 3.218(22)x10$^{-3}$ | 3.112(23)x10$^{-3}$ |
|  | $2P_\gamma^2(P_b^2-P_c^2)$ | C1 | $F_{bc}$ ($c_1$) | -0.0108(60)x10$^{-4}$ | -0.1955(97)x10$^{-4}$ |
|  | $(1-\cos 3\gamma)P^2$ | FV | $V_{3J}$ ($F_v$) | -2.3905(50)x10$^{-3}$ | -2.4324(69)x10$^{-3}$ |
|  | $(1-\cos 3\gamma)P_a^2$ | AK5 | $V_{3K}$ ($k_5$) | 1.114890(35)x10$^{-2}$ | 1.117844(23)x10$^{-2}$ |
|  | $(1-\cos 3\gamma)\{P_a,P_b\}$ | ODAB | $V_{3ab}$ ($d_{ab}$) | 8.9465(12)x10$^{-3}$ | 9.07791(65)x10$^{-3}$ |
|  | $(1-\cos 3\gamma)(P_b^2-P_c^2)$ | C2 | $V_{3bc}$ ($c_2$) | -8.574(21)x10$^{-5}$ | -8.698(21)x10$^{-5}$ |
|  | $\sin 3\gamma\{P_a,P_c\}$ | DAC | $D_{3ac}$ ($D_{ac}$) | 5.281(28)x10$^{-2}$ | 5.177(29)x10$^{-2}$ |
|  | $\sin 3\gamma\{P_b,P_c\}$ | DBC | $D_{3bc}$ ($D_{bc}$) | 0.7361(75)x10$^{-3}$ | 0.538(12)x10$^{-3}$ |
| {413} | $P_\gamma P_a P^2$ | ALV | $\rho_J$ ($L_v$) | -2.521(34)x10$^{-4}$ | -2.305(54)x10$^{-4}$ |
|  | $P_\gamma P_a^3$ | AK1 | $\rho_K$ ($k_1$) | -3.3807(16)x10$^{-2}$ | -3.4254(13)x10$^{-2}$ |
| | $P_\gamma(P_a^2 P_b + P_b P_a^2)$ | ODELTA | $\rho_{ab}$ ($\delta_{ab}$) | $4.749(33) \times 10^{-3}$ | $4.496(33) \times 10^{-3}$ |
| | $P_\gamma\{P_a,(P_b^2 - P_c^2)\}$ | C4 | $\rho_{bc}$ ($c_4$) | $-0.5055(58) \times 10^{-4}$ | $-0.7047(94) \times 10^{-4}$ |
| {404} | $-P^4$ | DJ | $\Delta_J$ | $1.625417(34) \times 10^{-6}$ | $1.688465(31) \times 10^{-6}$ |
| | $-P^2 P_a^2$ | DJK | $\Delta_{JK}$ | $10.74(17) \times 10^{-5}$ | $9.20(25) \times 10^{-5}$ |
| | $-P_a^4$ | DK | $\Delta_K$ | $8.195(13) \times 10^{-3}$ | $8.524(10) \times 10^{-3}$ |
| | $-2P^2(P_b^2 - P_c^2)$ | ODELN | $\delta_J$ | $5.5824(55) \times 10^{-8}$ | $5.9414(33) \times 10^{-8}$ |
| | $-\{P_a^2,(P_b^2 - P_c^2)\}$ | ODELK | $\delta_K$ | $5.584(11) \times 10^{-5}$ | $5.7361(89) \times 10^{-5}$ |
| | $\{P_a, P_b\} P^2$ | DABJ | $D_{abJ}$ | $0.124(23) \times 10^{-7}$ | $-0.548(23) \times 10^{-7}$ |
| | $\{P_a^3, P_b\}$ | DABK | $D_{abK}$ | $1.588(11) \times 10^{-3}$ | $1.443(11) \times 10^{-3}$ |
| {660} | $P_\gamma^6$ | AK4B | $F_{mm}$ ($k_{4B}$) | $0.7120(14) \times 10^{-5}$ | $1.01639(75) \times 10^{-5}$ |
| | $(1-\cos 9\gamma)/2$ | V9 | $V_9$ | $0.10518(67)$ | $-0.05126(34)$ |
| {651} | $P_\gamma^5 P_a$ | AK3B | $\rho_{mm}$ ($k_{3B}$) | $5.2689(65) \times 10^{-5}$ | $6.7042(35) \times 10^{-5}$ |
| {642} | $P_\gamma^4 P^2$ | AMV | $F_{mJ}$ ($M_v$) | $8.467(19) \times 10^{-8}$ | $9.215(14) \times 10^{-8}$ |
| | $P_\gamma^4 P_a^2$ | BK1 | $F_{mK}$ ($K_1$) | $1.5194(12) \times 10^{-4}$ | $1.79670(69) \times 10^{-4}$ |
| | $P_\gamma^4 \{P_a, P_b\}$ | DELTAB | $F_{mab}$ ($\Delta\Delta_{ab}$) | $0.237(28) \times 10^{-6}$ | $0.773(54) \times 10^{-6}$ |
| | $2P_\gamma^4(P_b^2 - P_c^2)$ | C3 | $F_{mbc}$ ($c_3$) | $0.092(12) \times 10^{-7}$ | $0.214(18) \times 10^{-7}$ |
| | $\{1-\cos 3\gamma, P_\gamma^2\} P^2$ | AK7J | $V_{3mJ}$ ($k_{7J}$) | $13.0(11) \times 10^{-6}$ | $9.4(16) \times 10^{-6}$ |
| | $(1-\cos 6\gamma) P^2$ | ANV | $V_{6J}$ ($N_v$) | $3.57(37) \times 10^{-5}$ | $2.64(53) \times 10^{-5}$ |
| | $(1-\cos 6\gamma) P_a^2$ | BK2 | $V_{6K}$ ($K_2$) | $-1.6233(58) \times 10^{-4}$ | $-1.3905(25) \times 10^{-4}$ |
| | $(1-\cos 6\gamma)\{P_a, P_b\}$ | ODAB6 | $V_{6ab}$ ($dd_{ab}$) | $-0.918(38) \times 10^{-4}$ | $-0.388(16) \times 10^{-4}$ |
| | $(1-\cos 6\gamma)(P_b^2 - P_c^2)$ | C11 | $V_{6bc}$ ($c_{11}$) | $-3.294(10) \times 10^{-5}$ | $-3.3840(70) \times 10^{-5}$ |
| | $\sin 6\gamma \{P_a, P_c\}$ | DAC6 | $D_{6ac}$ | $3.589(43) \times 10^{-4}$ | $3.401(58) \times 10^{-4}$ |
| {633} | $P_\gamma^3 P_a P^2$ | AK3J | $\rho_{mJ}$ ($k_{3J}$) | $7.940(68) \times 10^{-7}$ | $7.875(69) \times 10^{-7}$ |
| | $P_\gamma^3 P_a^3$ | AK3K | $\rho_{mK}$ ($k_{3K}$) | $2.2338(11) \times 10^{-4}$ | $2.51512(70) \times 10^{-4}$ |
| | $P_\gamma^3 \{P_a^2, P_b\}$ | ODELTB | $\rho_{mab}$ ($\delta\delta_{ab}$) | $0.486(57) \times 10^{-6}$ | $1.61(11) \times 10^{-6}$ |
| | $P_\gamma^3 \{P_a,(P_b^2 - P_c^2)\}$ | C12 | $\rho_{mbc}$ ($c_{12}$) | $7.594(68) \times 10^{-7}$ | $6.903(52) \times 10^{-7}$ |
| | $\{(1-\cos 3\gamma), P_a P^2 P_\gamma\}$ | AK6J | $\rho_{3J}$ ($k_{6J}$) | $3.61(17) \times 10^{-5}$ | $2.11(27) \times 10^{-5}$ |
| | $\{(1-\cos 3\gamma), P_a^3 P_\gamma\}$ | AK6K | $\rho_{3K}$ ($k_{6K}$) | $-1.558(79) \times 10^{-4}$ | $0.385(63) \times 10^{-4}$ |
| {624} | $P_\gamma^2 P^4$ | GVJ | $F_{JJ}$ ($g_v$) | $0.4941(23) \times 10^{-9}$ | $0.5243(22) \times 10^{-9}$ |
| | $P_\gamma^2 P_a^2 P^2$ | AK2J | $F_{JK}$ ($k_{2J}$) | $1.872(20) \times 10^{-6}$ | $1.769(19) \times 10^{-6}$ |
| | $P_\gamma^2 \{P_a, P_b\} P^2$ | DELTAJ | $F_{Jab}$ | $1.49(16) \times 10^{-9}$ | $2.75(21) \times 10^{-9}$ |
| | $2 P_\gamma^2 P^2 (P_b^2 - P_c^2)$ | C1J | $F_{Jbc}$ ($c_5$) | $-0.890(76) \times 10^{-9}$ | |
| | $P_\gamma^2 P_a^4$ | AK2K | $F_{KK}$ ($k_{2K}$) | $1.79004(55) \times 10^{-4}$ | $1.94907(47) \times 10^{-4}$ |
| | $P_\gamma^2 \{P_a^3, P_b\}$ | DELTAK | $F_{Kab}$ | $2.39(30) \times 10^{-7}$ | $8.29(59) \times 10^{-7}$ |
| | $P_\gamma^2 \{P_a^2, (P_b^2 - P_c^2)\}$ | C1K | $F_{Kbc}$ ($c_8$) | $1.650(20) \times 10^{-6}$ | $1.518(15) \times 10^{-6}$ |
| | $(1-\cos 3\gamma) P^4$ | OFV | $V_{3JJ}$ ($f_v$) | $8.590(23) \times 10^{-9}$ | $9.149(21) \times 10^{-9}$ |
| | $(1-\cos 3\gamma) P_a^2 P^2$ | AK5J | $V_{3JK}$ ($k_{5J}$) | $26.6(17) \times 10^{-6}$ | $7.6(26) \times 10^{-6}$ |
| | $(1-\cos 3\gamma)\{P_a, P_b\} P^2$ | ODABJ | $V_{3Jab}$ ($d_{abJ}$) | $-2.275(14) \times 10^{-7}$ | $-2.027(17) \times 10^{-7}$ |
| | $2(1-\cos 3\gamma)(P_b^2 - P_c^2) P^2$ | C2J | $V_{3Jbc}$ ($c_{2J}$) | $1.041(54) \times 10^{-9}$ | $1.251(43) \times 10^{-9}$ |
| | $(1-\cos 3\gamma) P_a^4$ | AK5K | $V_{3KK}$ ($f_k$) | $-2.37(13) \times 10^{-4}$ | $0.78(10) \times 10^{-4}$ |
| | $(1-\cos 3\gamma)\{P_a^3, P_b\}$ | ODABK | $V_{3Kab}$ ($d_{abK}$) | $-4.80(33) \times 10^{-7}$ | $-1.538(79) \times 10^{-7}$ |
| | $(1-\cos 3\gamma)\{P_a^2, (P_b^2 - P_c^2)\}$ | C2K | $V_{3Kbc}$ ($c_9$) | $7.667(83) \times 10^{-6}$ | $7.232(86) \times 10^{-6}$ |
| | $\sin 3\gamma P^2 \{P_a, P_c\}$ | DACJ | $D_{3acJ}$ | $-2.926(22) \times 10^{-7}$ | $-2.888(23) \times 10^{-7}$ |
| | $\sin 3\gamma P^2 \{P_b, P_c\}$ | DBCJ | $D_{3bcJ}$ | $-1.901(93) \times 10^{-8}$ | $-1.070(58) \times 10^{-8}$ |
| | $\sin 3\gamma \{P_a^3, P_c\}$ | DACK | $D_{3acK}$ | $-0.529(68) \times 10^{-6}$ | $0.70(10) \times 10^{-6}$ |
| | $\sin 3\gamma \{P_a^2, \{P_b, P_c\}\}$ | DBCK | $D_{3bcK}$ | $-1.348(63) \times 10^{-6}$ | $-0.585(70) \times 10^{-6}$ |
| {615} | $P_\gamma P_a P^4$ | OLV | $\rho_{JJ}$ ($l_v$) | $0.8090(68) \times 10^{-9}$ | $0.8961(62) \times 10^{-9}$ |
| | $P_\gamma P_a^3 P^2$ | AK1J | $\rho_{JK}$ ($\lambda_v$) | $1.351(15) \times 10^{-6}$ | $1.231(14) \times 10^{-6}$ |
| | $P_\gamma P^2 \{P_a^2, P_b\}$ | DAGJ | $\rho_{Jab}$ | $0.78(15) \times 10^{-9}$ | $1.91(18) \times 10^{-9}$ |
| | $P_\gamma P^2 \{P_a, (P_b^2 - P_c^2)\}$ | C4J | $\rho_{Jbc}$ ($c_7$) | $-0.32(12) \times 10^{-9}$ | $0.426(33) \times 10^{-9}$ |
| | $P_\gamma P_a^5$ | AK1K | $\rho_{KK}$ ($l_k$) | $7.5009(25) \times 10^{-5}$ | $7.9805(24) \times 10^{-5}$ |
| | $P_\gamma \{P_a^3, (P_b^2 - P_c^2)\}$ | C4K | $\rho_{Kbc}$ ($c_{7K}$) | $1.191(21) \times 10^{-6}$ | $1.119(16) \times 10^{-6}$ |
| {606} | $P^6$ | HJ | $H_J$ | $-1.194(21) \times 10^{-12}$ | $-1.191(16) \times 10^{-12}$ |
| | $P^4 P_a^2$ | HJK | $H_{JK}$ | $4.535(51) \times 10^{-10}$ | $4.781(40) \times 10^{-10}$ |
| | $P_a^4P^2$ | HKJ | $H_{KJ}$ | $2.653(49) \times 10^{-7}$ | $2.336(37) \times 10^{-7}$ |
| | $P_a^6$ | HK | $H_K$ | $1.29602(59) \times 10^{-5}$ | $1.35675(51) \times 10^{-5}$ |
| | $P^2\{P_a^2,(P_b^2-P_c^2)\}$ | OHJK | $h_{JK}$ | $0.538(54) \times 10^{-9}$ | $0.427(31) \times 10^{-9}$ |
| | $\{P_a^4,(P_b^2-P_c^2)\}$ | OHK | $h_K$ | $3.100(71) \times 10^{-7}$ | $2.928(54) \times 10^{-7}$ |
| {880} | $P_\gamma^8$ | AK4C | $F_{mmm}$ $(k_{4BB})$ | $0.5929(71) \times 10^{-7}$ | $-0.5887(30) \times 10^{-7}$ |
| {871} | $P_\gamma^7 P_a$ | AK3C | $\rho_{mmm}$ $(k_{3BB})$ | $0.4023(43) \times 10^{-6}$ | $-0.3447(19) \times 10^{-6}$ |
| {862} | $P_\gamma^6 P^2$ | AK4BJ | $F_{mmJ}$ | $-0.1511(81) \times 10^{-9}$ | $-0.4129(60) \times 10^{-9}$ |
| | $P_\gamma^6 P_a^2$ | AK4BK | $F_{mmK}$ | $1.166(11) \times 10^{-6}$ | $-0.8527(55) \times 10^{-6}$ |
| | $(1-\cos 9\gamma)P^2$ | V9J | $V_{9J}$ | $13.71(32) \times 10^{-6}$ | $-1.31(66) \times 10^{-6}$ |
| | $(1-\cos 9\gamma)\{P_a,P_b\}$ | ODAB9 | $V_{9ab}$ | $1.50(11) \times 10^{-4}$ | $-0.819(43) \times 10^{-4}$ |
| {853} | $P_\gamma^5 P_a P^2$ | AK3BJ | $\rho_{mmJ}$ | $-0.710(26) \times 10^{-9}$ | $-1.635(26) \times 10^{-9}$ |
| | $P_\gamma^5 P_a^3$ | AK3BK | $\rho_{mmK}$ | $1.869(16) \times 10^{-6}$ | $-1.1548(88) \times 10^{-6}$ |
| {844} | $P_\gamma^4 P_a^2 P^2$ | G4J2K2 | $F_{mJK}$ $(K_{1J})$ | $-0.992(27) \times 10^{-9}$ | $-2.097(63) \times 10^{-9}$ |
| | $P_\gamma^4 P^2\{P_a,P_b\}$ | DG4J | $F_{mJab}$ | | $-8.67(73) \times 10^{-12}$ |
| | $P_\gamma^4 P_a^4$ | G4K4 | $F_{mKK}$ $(K_{1K})$ | $1.789(13) \times 10^{-6}$ | $-0.9220(84) \times 10^{-6}$ |
| | $P_\gamma^4\{P_a^2,(P_b^2-P_c^2)\}$ | G4BCK | $F_{mKbc}$ | | $0.886(95) \times 10^{-10}$ |
| | $(1-\cos 6\gamma)P^4$ | C6J4 | $V_{6JJ}$ $(N_{vJ})$ | | $4.44(32) \times 10^{-10}$ |
| | $(1-\cos 6\gamma)P^2 P_a^2$ | C6J2K2 | $V_{6JK}$ | $2.409(52) \times 10^{-7}$ | $1.953(61) \times 10^{-7}$ |
| | $(1-\cos 6\gamma)P^2\{P_a,P_b\}$ | CABJ | $V_{6Jab}$ | $5.43(18) \times 10^{-8}$ | $3.50(15) \times 10^{-8}$ |
| | $2(1-\cos 6\gamma)P^2(P_b^2-P_c^2)$ | C6BCJ | $V_{6Jbc}$ | $0.625(72) \times 10^{-9}$ | $1.326(52) \times 10^{-9}$ |
| | $(1-\cos 6\gamma)P_a^4$ | C6K4 | $V_{6KK}$ | $-1.96(15) \times 10^{-7}$ | $-3.143(66) \times 10^{-7}$ |
| | $(1-\cos 6\gamma)\{P_a^3,P_b\}$ | CABK | $V_{6Kab}$ $(dd_{abK})$ | | $2.26(21) \times 10^{-7}$ |
| | $(1-\cos 6\gamma)\{P_a^2,(P_b^2-P_c^2)\}$ | C6BCK | $V_{6Kbc}$ $(c_{11K})$ | $-1.014(33) \times 10^{-7}$ | $-1.351(43) \times 10^{-7}$ |
| {835} | $P_\gamma^3 P_a^3 P^2$ | GAJ2K2 | $\rho_{mJK}$ | $-0.476(12) \times 10^{-9}$ | $-1.062(64) \times 10^{-9}$ |
| | $P_\gamma^3 P_a^5$ | GAK4 | $\rho_{mKK}$ | $10.226(70) \times 10^{-7}$ | $-4.305(48) \times 10^{-7}$ |
| | $P_\gamma^3\{P_a^3,(P_b^2-P_c^2)\}$ | AG3BCK | $\rho_{mKbc}$ | | $8.66(94) \times 10^{-11}$ |
| | $\{1-\cos 3\gamma, P_a^3 P_\gamma\}P^2$ | AK6JK | $\rho_{3JK}$ | $6.56(30) \times 10^{-8}$ | $5.58(22) \times 10^{-8}$ |
| {826} | $P_\gamma^2 P_a^2 P^4$ | GJ4K2 | $F_{JJK}$ | | $1.477(32) \times 10^{-12}$ |
| | $P_\gamma^2 P_a^4 P^2$ | GJ2K4 | $F_{JKK}$ | | $-1.90(20) \times 10^{-10}$ |
| | $P_\gamma^2 P^2 \{P_a^3, P_b\}$ | DELTJK | $F_{JKab}$ | | $4.53(59) \times 10^{-12}$ |
| | $P_\gamma^2 P_a^6$ | GK6 | $F_{KKK}$ | $3.240(20) \times 10^{-7}$ | $-1.068(15) \times 10^{-7}$ |
| | $P_\gamma^2 \{P_a^5, P_b\}$ | DELTKK | $F_{KKab}$ | | $-0.89(11) \times 10^{-11}$ |
| | $(1-\cos 3\gamma) P^2 P_a^4$ | FJ2K4 | $V_{3JKK}$ | $1.072(48) \times 10^{-7}$ | $0.914(36) \times 10^{-7}$ |
| | $(1-\cos 3\gamma) P_a^6$ | FK6 | $V_{3KKK}$ | | $-1.00(46) \times 10^{-10}$ |
| | $\sin 3\gamma P^4 \{P_a, P_c\}$ | DACJJ | $D_{3acJJ}$ | | $-0.717(46) \times 10^{-11}$ |
| | $\sin 3\gamma P^2 \{P_a^3, P_c\}$ | DACJK | $D_{3acJK}$ | $-1.323(35) \times 10^{-9}$ | $-1.593(30) \times 10^{-9}$ |
| {817} | $P_\gamma P_a^3 P^4$ | AGJ4K2 | $\rho_{JJK}$ | | $1.192(20) \times 10^{-12}$ |
| | $P_\gamma P_a^5 P^2$ | AGJ2K4 | $\rho_{JKK}$ | $0.396(19) \times 10^{-10}$ | |
| | $P_\gamma P_a^7$ | AGK6 | $\rho_{KKK}$ ($l_{KK}$) | $4.404(26) \times 10^{-8}$ | $-1.033(21) \times 10^{-8}$ |
| {10 10 0} | $P_\gamma^{10}$ | AK4D | $F_{mmmm}$ | | $-0.940(20) \times 10^{-10}$ |
| {10 9 1} | $P_\gamma^9 P_a$ | AK3D | $\rho_{mmmm}$ | | $-4.663(92) \times 10^{-10}$ |
| {10 8 2} | $P_\gamma^8 P_a^2$ | AK4CK | $F_{mmmK}$ | | $-0.939(17) \times 10^{-9}$ |
| {10 7 3} | $P_\gamma^7 P_a^3$ | AK3CK | $\rho_{mmmK}$ | | $-0.957(17) \times 10^{-9}$ |
| {10 6 4} | $P_\gamma^6 P_a^4$ | AK4BK4 | $F_{mmKK}$ | | $-4.915(82) \times 10^{-10}$ |
| {10 5 5} | $P_\gamma^5 P_a^5$ | AK3BK4 | $\rho_{mmKK}$ | | $-1.017(16) \times 10^{-10}$ |

[a] Order of the Hamiltonian term in the notation of Ref. [31]: $n = l + m$, where $n$ is the total order of the operator, $l$ is the order of the torsional factor, and $m$ is the order of the rotational factor. Note that the rotational-order $m$ of Ref. [31], used in the first column of this table, does not have the same meaning as the subscript $m$ of Ref. [18], used in the fourth column of this table.

[b] $\{A,B\} \equiv AB + BA$. The product of the parameter and operator from a given row yields the term actually used in the torsion-rotation Hamiltonian of the fitting program, except for F, $\rho$ and A, which occur in the Hamiltonian in the form $F(P_\gamma + \rho P_a)^2 + A P_a^2$.

[c]The parameter labels given in this column are the ones used in this manuscript. The labels in parentheses have been employed in earlier literature.

[d]Parameter uncertainties are given in parentheses, and represent one standard deviation in the last digit (type A, k=1, [32]).

[e]$\rho$ is unitless, but all subscripted versions of $\rho$ (e.g., $\rho_m$, $\rho_J$, $\rho_K$, etc.) are in cm$^{-1}$.

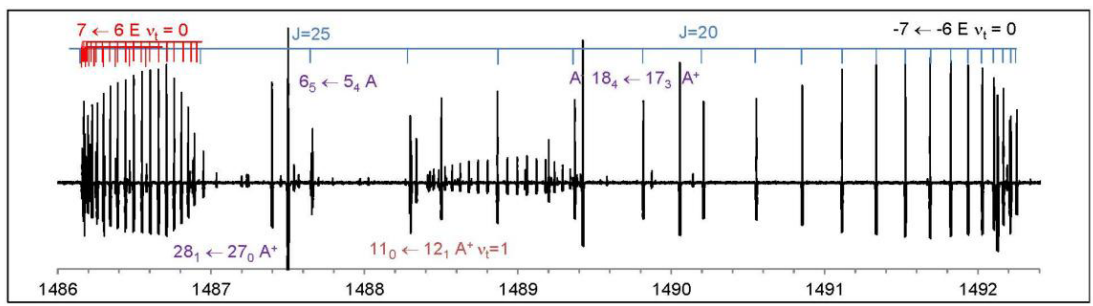

Fig.1 Section of the THz spectrum of $^{13}CH_3OH$ illustrating the $K = 7 \leftarrow 6$ and $-7 \leftarrow -6$ $v_t = 0$ $Q$ branches of $E$ torsional symmetry. Several other strong $v_t = 0$ transitions are also seen, labelled in $J_K$ notation, plus one $v_t = 1$ line. The smaller $Q$ branch in the centre of the picture so far is unidentified.

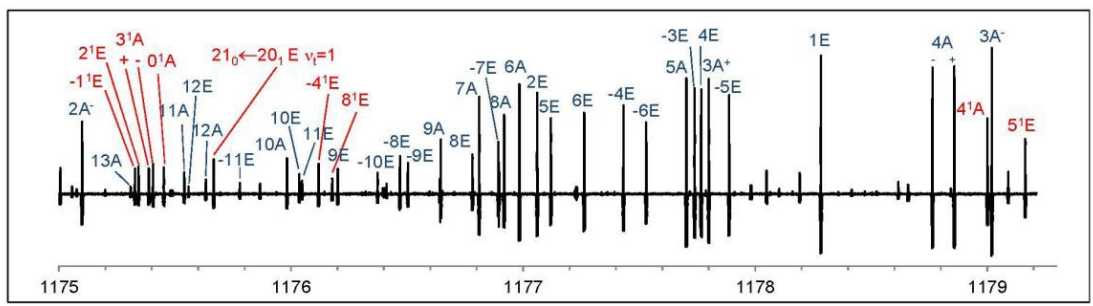

Fig.2   Part of the dense $J_K = 25_K \leftarrow 24_K$ $a$-type $J$-multiplet in the THz spectrum of $^{13}$CH$_3$OH. Transitions are labelled with the $K$ value and the torsional $A$ or $E$ symmetry; $v_t = 0$ lines (labelled in blue) have no superscript on $K$ while $v_t = 1$ lines (labelled in red) have superscript 1. The $J_K = 21_0 \leftarrow 20_1$ $E$ $v_t = 1$ $b$-type line (labelled in pink) is also seen.